\documentclass[letterpaper,twocolumn,10pt]{article}
\usepackage{usenix-2020-09}

\usepackage{tikz}
\usepackage{amsmath}
\usepackage{url}
\usepackage{hyperref}
\usepackage{paralist}
\usepackage{titling}

\newcommand{\mirage}{\textsf{MIRAGE}}

\newcommand{\sass}{\textsf{SassCache}}

\definecolor{lightgray}{HTML}{F7F7F7}
\definecolor{customblack}{HTML}{A0A0A0}
\title{Addendum: Systematic Evaluation of Randomized Cache Designs against Cache Occupancy \\[1ex]
{\large Presented at USENIX Security 2025}}

\author{%
Anirban Chakraborty$^{\dagger}$ \and
Nimish Mishra$^{\ddagger}$ \and
Sayandeep Saha$^{\S}$ \and
Sarani Bhattacharya$^{\ddagger}$ \and
Debdeep Mukhopadhyay$^{\ddagger}$\\[1ex]
$^{\dagger}$Max Planck Institute for Security and Privacy, Germany\\
$^{\ddagger}$Indian Institute of Technology Kharagpur, India\\
$^{\S}$Indian Institute of Technology Bombay, India\\[1ex]
\footnotesize
\href{mailto:anirban.chakraborty@mpi-sp.org}{anirban.chakraborty@mpi-sp.org}~~
\href{mailto:nimish.mishra@kgpian.iitkgp.ac.in}{nimish.mishra@kgpian.iitkgp.ac.in}~~
\href{mailto:debdeep@cse.iitkgp.ac.in}{\{sarani, debdeep\}@cse.iitkgp.ac.in}~~
\href{mailto:sayandeepsaha@cse.iitb.ac.in}{sayandeepsaha@cse.iitb.ac.in}
}

\date{}

\begin{document}
\maketitle

In the main text published at USENIX Security 2025~\cite{chakraborty2025systematic}, we presented a systematic analysis of the role of cache occupancy in the design considerations for randomized caches (from the perspectives of performance and security). On the performance front, we presented a uniform benchmarking strategy that allows for a fair comparison among different randomized cache designs. Likewise, from the security perspective, we presented three threat assumptions: (1) covert channels; (2) process fingerprinting side-channel; and (3) AES key recovery. The main takeaway of our work is an open problem of designing a randomized cache of comparable efficiency
with modern set-associative LLCs, while still resisting both
contention-based and occupancy-based attacks. 

This note is meant as an addendum to the main text in light of the observations made in~\cite{cao2025yet}. To summarize, the authors in~\cite{cao2025yet} argue that (1) L1d cache size plays a role in adversarial success; and that (2) a patched version of \mirage~with randomized initial seeding of global eviction map prevents leakage of AES key. We discuss the same in this addendum.

\section{Sources Used, Attack Reproducibility, and Number of Traces needed in~\cite{chakraborty2025systematic}}

In~\cite{chakraborty2025systematic}, we used the sources of \mirage~available to us at the time of writing\footnote{\url{https://github.com/gururaj-s/mirage}; commit: \texttt{2c763da}} to develop a downstream artifact for our evaluation\footnote{\url{https://github.com/SEAL-IIT-KGP/randomized_caches}}. We maintain that the AES key recovery described in~\cite{chakraborty2025systematic} is reproducible on those original sources, and has also been likewise reproduced in revision 3 of~\cite{cao2025yet} (in version $1$, the authors reported the inability to reproduce our results out-of-the-box). 

Moreover, the exact number of traces needed to observe a drop in Guessing Entropy will differ across runs of the attack. This is an expected consequence of using a statistical attack, as we do in~\cite{chakraborty2025systematic}. Limited number of side-channel traces (as is the case with any real world attack) approximate the actual underlying distribution to different statistical distances (more traces imply better approximation), and thus these approximations vary across repeated runs of the attack. Therefore, while the \textit{exact} number of traces needed to achieve the guessing entropy of $32$ (or below) will vary, we maintain the \textit{overall general trend} of reduction in guessing entropy for \mirage~remains intact when the attack is executed on the original sources available to us at the time of writing.

\section{Role of L1d cache}

A point raised in~\cite{cao2025yet} is the role of L1d in the attack: an implication that the original attack in~\cite{chakraborty2025systematic} was successful due to its unrealistic setting of L1d cache size, and that using a larger L1d cache size (ex. $64$ KB) masks the leakage. We first note the role the L1d cache plays in the attack and then provide clarifications on the chosen L1d size. Since the L1 instruction cache is separate from L1d, it bears no consequence to the attack; we thus omit it in our discussion here. We begin by first recalling the attack overview from Sec.7.1 in~\cite{chakraborty2025systematic}:

\begin{enumerate} \itemsep0em
    \item The attacker first lets the AES victim setup its secret key, and precompute T-Tables.

    \item Before the experiment begins, the attacker fills the LLC with spurious occupancy. These memory accesses are \textit{never} re-accessed again during the course of the actual attack. This step also ensures that the AES victim T-Tables are reliably flushed from the LLC.

    \item Given a fixed occupancy $X\%$, the attacker \texttt{malloc}s about $\frac{16\times X}{100}$ MB of memory, and repeatedly accesses it to ensure occupancy of complete L1d and $X\%$ of LLC, since LLC is inclusive cache\footnote{Most cache designs (including the ones we evaluate here) consider an inclusive cache-hierarchy.}.

    \item Given a randomly generated plaintext $P$, the attacker then lets AES victim run a \textit{single} encryption of $P$, and obtains the ciphertext $C$.

    \item Finally, the attacker uses \texttt{rdtsc} to time access to its previously allocated $\frac{16\times X}{100}$ MB of memory. Call it $T$.
\end{enumerate}

Note here that in both step (2), the adversary needs to cover the entirety of L1d cache. This is intentional: in step (1), when the T-tables are initialized, they reside in the L1d cache. If the attack is carried out without flushing the T-table (i.e. execute step (3) directly after step (1), skipping step (2)), the \textbf{attack would still work}, but the leakage source would then be contention in the L1d cache, and not the LLC. Conclusions from such an attack would be of no consequence to the LLC, which is where the randomized mappings studied in~\cite{chakraborty2025systematic} function.

It is therefore imperative that the tables are flushed from the cache hierarchy post initialization to allow us to study effects of LLC cache line installs on different randomized cache designs. Below is detailed a lifecycle of a single T-table entry (abbr. $\mathbf{t}$) throughout the different stages of our attack (extension to the entire T-table is straightforward):

\begin{itemize}
    \item Entry $\mathbf{t}$ is initialized by the AES victim. Since this initialization uses \texttt{load} and \texttt{store} instructions, the entry $\mathbf{t}$ occupies a cache line in the L1d cache.

    \item Since we do not assume adversarial capability to use ISA (like \texttt{clflush} on \texttt{x86} ISA) to flush $\mathbf{t}$, we flush the \textit{entire} L1d and the \textit{entire} LLC to ensure $\mathbf{t}$ is reliably flushed from the LLC.

    \item When the victim makes the first T-table access $\mathbf{t}$ during AES encryption, because of Step 2, we are guaranteed to have a \textit{cache line install} in the LLC. For the specific case of \mirage, we are guaranteed to have a global eviction with all but negligible probability. Post the cache line install in LLC, $\mathbf{t}$ is installed in the L1d, and utilized by the AES victim process.
\end{itemize}

In order to study contention in the presence of adversarial cache occupancy in the LLC, the event of interest (and subsequent leakage source exploited in~\cite{chakraborty2025systematic}) is the event of cache line install in the LLC. \textbf{There is no role of L1d in the attack, since the exploited leakage source across different randomized cache designs is the manner of cache line installs in the LLC, and effect of such LLC line installs on adversarial occupancy of the LLC}. The fact that such LLC line installs transitively have L1d line installs is a consequence of an inclusive cache hierarchy, and was not the source of the leakage exploited in~\cite{chakraborty2025systematic}.

\subsection{Choice of L1d size in~\cite{chakraborty2025systematic}}

The choice of L1d size in~\cite{chakraborty2025systematic} was thereby driven by the need to \textit{minimize the gem5 simulation time}, in order to not spend simulation cycles on a micro-architectural element (i.e. L1d cache) not being targeted by the attack. Tab.~\ref{tab:tick_comparison} gives a comparative of the gem5 simulation ticks; observe the additional work \textit{per} AES execution being performed by the simulator, which scales linearly with the attack complexity \textit{per} cache design. We stress that such a consideration on simulation time is solely for the platform available to us for testing (gem5) and does not apply to a (future) implementation of randomized caches on real hardware, as also acknowledged in the artifact appendix accompanying~\cite{chakraborty2025systematic}:

\vspace{3mm}
\textit{We encourage the user to try out data collection with different keys to get a trend (and GE) closer to what is reported in the paper. However, as also noted in the paper (footnote $21$), the rate of data collection is at best $500$ observations per hour. We were able to thereby deploy about $350$ cores per Intel Xeon server, across three such servers. The overall data collection for \textit{all} designs and multiple keys took over 2 weeks of compute hours. [...]}

\textit{Note that such an inhibitory rate is \textit{not a problem of attack design}, but rather is the consequence of the gem5 simulations (which is the go-to simulation strategy for state-of-the-art randomized cache literature). In a realistic setting (when these randomized caches are deployed on real hardware), our attack will be much faster.
}

\vspace{3mm}

\begin{figure}[!t]
    \centering
    \includegraphics[scale=0.6]{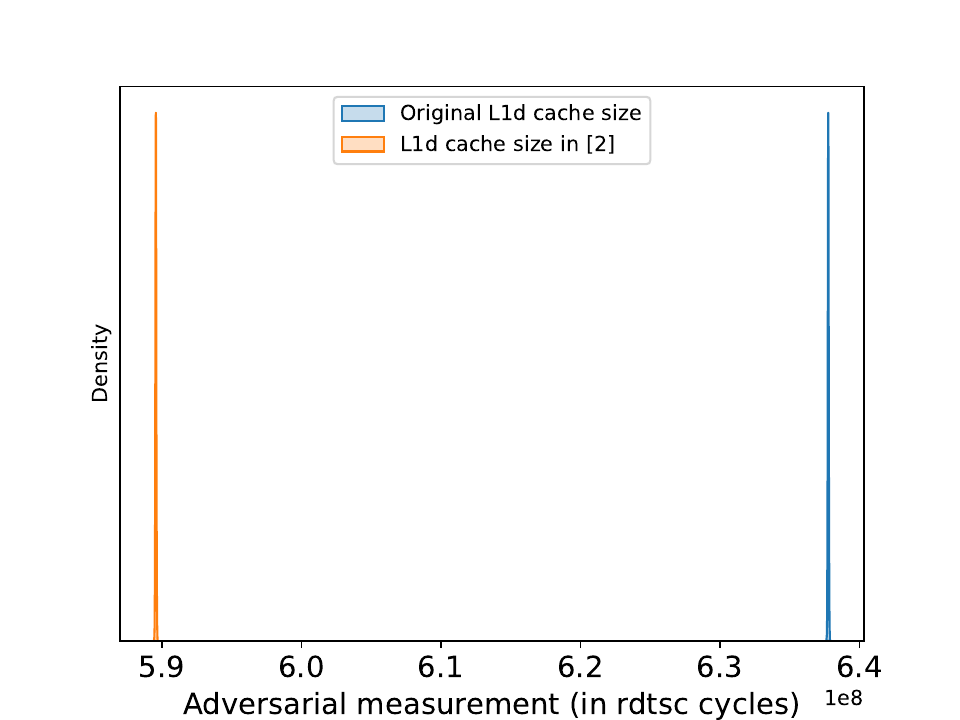}
    %\vspace{-3mm}
    \caption{Kernel Density Estimators for (1) the original L1d cache size considered in~\cite{chakraborty2025systematic}; vs (2) the L1d cache size in~\cite{cao2025yet}. The shape assumed by both distribution is gaussian.}
    %\vspace{-4mm}
    \label{fig:l1d_comparison}
\end{figure}

In Fig.~\ref{fig:l1d_comparison}, we compare the Kernel Density Estimators (KDEs) of the adversarial measurements of its own occupancy (Step $5$ in the attack overview detailed earlier this section) for the considered L1d cache size in~\cite{chakraborty2025systematic} vs the size used in~\cite{cao2025yet}. Notice both distributions have the same gaussian envelope with different means, with the larger L1d cache having an expected lower mean. In other words, the statistical attack in~\cite{chakraborty2025systematic} exploits the difference in the (gaussian) distribution approximation between two attack runs (first with a profiled key; second with the victim key). Changing the L1d cache relatively displaces \textit{both} distributions in the X-axis (wrt. adversarial timing measurement) but does not distort the gaussian envelope itself, as clear from Fig.~\ref{fig:l1d_comparison}.

Finally, on the original \mirage~sources considered in~\cite{chakraborty2025systematic} with the L1d cache size considered in~\cite{cao2025yet}, \textit{we were able to observe AES key leakage}. \textbf{We maintain our position thereby: on the sources available to us at the time of writing, increasing L1d cache size does not prevent the leakage introduced by global evictions during LLC line installs.}

\begin{table}[!t]
\centering
\resizebox{0.4\textwidth}{!}{
\begin{tabular}{|c|c|c|}
\hline
\mirage~Configuration & L1d size & Simulation ticks\\
\hline
\hline
Original sources & Original Paper & $837793997703$ \\
\hline
Original sources & Same as~\cite{cao2025yet} & $881671966812$ \\
\hline
\end{tabular}
}
\caption{Number of recorded gem5 ticks for a single AES execution.}
\label{tab:tick_comparison}
\end{table}

\section{Role of seeding global eviction mapping}

Another point raised in~\cite{cao2025yet} was the release of a new patch which randomizes the seeding of the global eviction mapping, hereafter abbreviated as~\mirage+ to indicate (1) the original sources available to us at the time of writing with (2) L1d cache size in~\cite{cao2025yet}, plus the (3) additional patch of randomizing the global eviction mapping. The observation in~\cite{cao2025yet} was that~\mirage+ prevents the AES leakage exploited in~\cite{chakraborty2025systematic}.

On our end, we built \mirage+ from source\footnote{\url{github.com/sith-lab/yet-another-mirage-of-breaking-mirage}}, and applied the provided patches as recommended in~\cite{cao2025yet}. As a first level of analysis, we chose to perform the following experiment:

\begin{itemize}
    \item \textbf{Run 1}: Fix plaintext input \texttt{0xa7d960e3eac4b884fdcde51438edb007} borrowed from~\cite{cao2025yet} and AES key \texttt{0x7766554433221100ffeeddccbbaa9988} (originally considered in~\cite{chakraborty2025systematic}). Collect traces using the script \texttt{run\_mirage\_fixedCT.sh} provided in~\cite{cao2025yet}. Call these traces $T_{1}$.

    \item \textbf{Run 2:} Similar as run 1, except the victim key is now changed to \texttt{0xffeeddccbbaa99887766554433221100} (originally considered in~\cite{chakraborty2025systematic}). Call such traces $T_{2}$.

    \item \textbf{Test:} Apply the standard minimum-maximum scaler\footnote{\url{https://scikit-learn.org/stable/modules/generated/sklearn.preprocessing.MinMaxScaler.html}} on the data, and perform Welch's T-test~\cite{becker2013test} on $T_{1}$ and $T_{2}$ to determine whether the distributions assumed by the two runs are significantly different.
\end{itemize}

This experiment serves as a first level analysis of whether \mirage+~leaks, with such testing having been a literature standard for side-channel leakage evaluations. In case the test fails to provide evidence of a leakage, further evaluations (like the statistical attack in~\cite{chakraborty2025systematic}) is fruitless effort. 

Our test results gave a \textbf{T-statistic = 7.1269; pvalue = 1.42e-12}. A negligibly small p-value indicates that, if the null hypothesis were true, the likelihood of observing such a difference is negligible low; and an above-threshold T-statistic establishes leakage. In simpler terms, the interpretation of the test is that $T_{1}$ and $T_{2}$ are statistically different. Recall that the \textit{only} difference between $T_{1}$ and $T_{2}$ is the secret AES key, further solidifying that \mirage+ leaks statistically by virtue of using different AES keys. A full key recovery attack will follow similar lifecycle as in~\cite{chakraborty2025systematic}, with the additional recommendation of including (1) \textit{noise averaging}\footnote{See~\cite{standaert2018not} for examples where noise averaging improves Signal-to-Noise ratio (SNR) and reduces side-channel attack complexity in terms of number of traces needed.}; and (2) gaussian tail-cutting, to offset the \texttt{rdtsc} cycle variations observed in~\cite{cao2025yet}.

Therefore, \textbf{we maintain our position that, as noted in an observation in~\cite{cao2025yet}, (1) increased L1d cache size; and (2) randomizing global eviction mapping, are \textit{not} self-sufficient to \textit{prevent} AES leakage} (see next section on our takeaway from this). At the same time, we note that an investigation of the attack parameters of full key recovery on a design unavailable to us at the time of writing~\cite{chakraborty2025systematic} is beyond the scope of this addendum, and is left for a follow-up work.

\section{Discussion and Takeaways}

From a broader perspective, the initial exploit in~\cite{chakraborty2025systematic} and the leakage in \mirage+ is the consequence of a fundamental facet of (randomized) cache design: allowing two processes to \textit{contend} for the same hardware. As such, while the exact attack parameters differ, we find it unsurprising that \textit{contention} leads to \textit{observable leakage} dependent on secret cryptographic material. 

\vspace{1mm}

We reassert our conclusions from~\cite{chakraborty2025systematic}: the way forward to prevent a side-channel attack as ours is to avoid the fundamental leakage source itself- contention in LLC. As noted in the main paper too, \sass~\cite{giner2022scatter} does this through its domain isolation, and we observed it to exhibit the highest resilience to all our attacks. Although compartmentalization nevertheless incurs performance overheads, we believe it is the correct strategy from a security perspective since it aims to eliminate the root cause of LLC contention itself, which other designs included in~\cite{chakraborty2025systematic} (as well as \mirage+ as discussed in this addendum) fundamentally do not allow.

\bibliographystyle{plain}
\bibliography{refs}

%%%%%%%%%%%%%%%%%%%%%%%%%%%%%%%%%%%%%%%%%%%%%%%%%%%%%%%%%%%%%%%%%%%%%%%%%%%%%%%%
\end{document}